\documentclass[12pt]{article}

\usepackage{newtxtext,newtxmath}

\usepackage{graphicx}

\usepackage[letterpaper,margin=1in]{geometry}

\renewenvironment{abstract}
	{\quotation}
	{\endquotation}

\date{}

\makeatletter
\renewcommand{\fnum@figure}{\textbf{Figure \thefigure}}
\renewcommand{\fnum@table}{\textbf{Table \thetable}}
\makeatother

\usepackage{scicite}

\usepackage{url}

\def\scititle{
	Reducing Catastrophic Risk from AI with Systematic Monitoring and Evaluation of Rogue AI Progression
}
\title{\bfseries \boldmath \scititle}

\author{
	T. Bauer$^{1}$${^\ast}$,\and
        W.P. Kegelmeyer$^{2}$,\and
    E. Begoli$^{3}$,\and 
    A. Sadovnik$^{3}$,\and 
    T. Emerson$^{4}$,\and 
    C. Corley$^{4}$,\and 
    N. Generous$^{5}$,\and
    J. Moore$^{5}$,\and
    B. Bartoldson$^{6}$,\and 
    R. Goldhahn$^{6}$,\and 
    M. Goldman$^{6}$,\and 
    M. Greaves$^{7}$,\and
    M. J. D. Vermeer$^{8}$,\and
    B. MacLennan$^{9}$,\and
    D. Schulker$^{10}$,\and
    N. VanHoudnos$^{10}$,\and
    J. Bansemer$^{11}$,\and
    Y. Bengio$^{12}$
    \\
	\small$^{1}$Sandia National Laboratories (SNL), Albuquerque, NM, 87185, USA.\and
	\small$^{2}$Sandia National Laboratories, Livermore, CA 94451, USA.\and
    \small$^{3}$Oakridge National Laboratories (ORNL), Oak Ridge, TN, 37830, USA.\and
    \small$^{4}$Pacific Northwest National Laboratory (PNNL), Richland, WA, 99352, USA.\and 
    \small$^{5}$Los Alamos National Laboratory, Los Alamos, NM, 87545, USA.\and
    \small$^{6}$Lawrence Livermore National Labs (LLNL), Livermore, CA, 94550, USA.\and
    \small$^{7}$Schmidt Sciences, New York, NY, 10011, USA.\and
    \small$^{8}$RAND, Santa Monica, CA 90401, USA.\and
    \small$^{9}$University of Tennessee, Knoxville, TN 37996, USA.\and
    \small$^{10}$Software Engineering Institute (SEI), Carnegie Mellon University, Pittsburgh, PA, 15213, USA.\and
    \small$^{11}$Georgetown University, Washington D.C., 20057, USA.\and
    \small$^{12}$University of Montreal and LawZero, Montreal, H2S 3G9, Canada.\and
	\small$^\ast$Corresponding author. Email: tlbauer@sandia.gov\and
}

\begin{document} 

\maketitle

\vspace{5mm}
\begin{abstract} \bfseries \boldmath
This article presents a structured framework of behavioral indicators that may signal progression toward potentially catastrophic threats from artificial intelligence systems. 
We adopt a pragmatic approach, inspired by established methodologies in cybersecurity and 
national security.  
By establishing clear metrics, indicators, and thresholds across multiple dimensions of AI capability and behavior, this framework enables researchers and policymakers to implement evidence-based monitoring protocols.
\end{abstract}

On January 7 and 8, 2025, several Federally Funded Research and Development Centers (FFRDCs) organized a small, focused, cross‑lab workshop to consider how to assess and monitor the possibility that rogue AI could present an existential threat. There were two attendees from each of the participating labs (Sandia National Laboratories, Pacific Northwest National Laboratory, Los Alamos National Laboratory, Lawrence Livermore National Laboratory, Oak Ridge National Laboratory, Software Engineering Institute) and eleven notable external invitees, including Yoshua Bengio of the University of Montreal, Melanie Mitchell of the Santa Fe Institute, and John Bansemer of CSET. The discussions were hosted at LANL and facilitated by Sandia business staff. Over the following few months, the participants collaborated to write this summary of our discussions.

As AI systems become more autonomous, they may act outside human oversight, especially as capabilities grow. A powerful, uncontrolled AI would pose a new threat universe beyond current security doctrines, which are driven by human motives. Such AI could render traditional deterrence ineffective, creating a significant strategic vulnerability.

By taking a framework-based approach, we develop a way of considering this problem independently of the debate regarding the likelihood of this threat's realization. We consider the conditions necessary for the realization of this type of risk and suggest \textit{observables} that can be monitored.


Our framework consists of five categories defined by specific AI behaviors and related indicators and observables. These categories, listed in Table~\ref{tbl:categories}, can also be roughly thought of as stages in that the latter categories presuppose the existence of the earlier ones. For each category, we will provide a definition, followed by example observables and indicators. An ``observable'' in our framework is a measurable or detectable attribute, event, or property in an environment without any associated judgment. For example, whether an AI system can persist data, and for how long, is an observable. An indicator is a particular value, pattern, or instance of an observable that may raise concerns. An example of an indicator would be an AI system autonomously using steganography to hide information in persistent storage.


\textbf{Category 1: Motivation:}
An AI system whose motivations match the intentions of the developers is considered to be ``aligned,'' and otherwise ``misaligned'' \cite{ngo2022alignment}. There are two primary ways in which well-intentioned specifications could result in AI systems with misaligned goals: \textbf{misspecification} \cite{paneffects} or \textbf{misgeneralization} \cite{shah2022goal}.

When a design is misspecified, an AI may take actions that are correct under the specification but are against the intent of the developer. In the context of reinforcement learning, misspecification manifests as ``reward hacking'' \cite{skalse2022defining}, where AI systems score highly on the specified reward function, while somehow ``gaming'' the system with respect to the developer's desires.

Goal misgeneralization occurs when an AI learns a set of goals that are compatible with the specification during training or development but which generalize in undesired ways during deployment. Notably, this can occur even when goals are properly specified or when an AI is not actively exploiting any misspecifications and may pose a greater concern in tandem.

If an AI system's motivations are not explicitly stated, they may be inferred from the system's actions, i.e., its revealed motivations. The sum of the model's actions across all deployments should inform our assessments of its motivations, though certain \textit{observables} will be more relevant and concerning than others.

The key indicators for misalignment are AI systems that reason and plan explicitly about motivations in conflict with their developers' intentions, and systems that exceed the bounds of normal operation in egregious or consistent ways for unclear reasons. These indicators, when paired with the capability to escape human control, pose significant risks.

Specific actions may be hard to predict in advance, but indicators of notable concern would include, for example, explicitly expressing misaligned motivations, especially those related to escaping human control or causing harm, and acting deceptively to avoid or disable oversight mechanisms.


\textbf{Category 2: Persistence:}
By ``persistence,'' we mean an AI system's capacity to maintain and influence its state over an extended period of time and space. Without the ability to project itself persistently across time, an AI system cannot engage in sophisticated behavior or experimentation.

AI systems have already exceeded the persistence threshold required to pose a major risk. LLM‑based AI systems have transitioned from text‑only chatbots to agentic AI systems that have access to database systems in the real world.

Monitoring persistence in AI systems is important for accurate risk assessments. The scale of persistence, especially in key risk areas, impacts the ease with which an AI system could cause harm. There is significant commercial incentive to continue developing increasingly sophisticated large scale persistence mechanisms.  Effective monitoring, alignment, and safety controls can limit the danger posed by increasingly sophisticated systems.

The key observable for persistence is whether and how ephemeral an AI's data is. Some persistence is highly ephemeral, only lasting a short amount of time. One example of this is context. Chain of Thought protocols in LLMs (Large Language Models), for example, generate text, store it in the context, and refer to that later in the course of responding to a single query. AI systems designed to operate autonomously for extended periods of time \cite{kwa2025measuring} may be able to keep context for longer durations. AI with access to tooling may have access to databases that persist for years, or the ability to write web pages that ultimately get archived in the Internet Archive and live on indefinitely.  Also consider agents in contemporary coding assistant systems. An ``agent'' may be a record of a single conversation between a coder and an AI. This conversation can persist indefinitely within the integrated development environment, allowing the user to return to it later. This effectively provides long-term storage for the AI system, which can also modify code and often execute commands on the machine where it resides.

The key indicator of rogue behavior in an AI is whether the data storage used is in alignment with how the system is intended to be used. In the simplest case, this can be tested directly, but there are various ways misaligned data can be detected. One such method has to do with information density. A misaligned AI needs the ability to persist information that is not necessary for the goals of the user. If it must be deceptive, it may have to store the information hidden within the data needed for legitimate uses. In such cases, it may be necessary to use something like steganography to hide in the open. Recent research has also uncovered other more subtle forms of persistence where biases may be indirectly invoked through seemingly unrelated data~\cite{cloud2025subliminallearninglanguagemodels}. In a situation like this, the persistence might be a tendency to move future models or fine‑tuned models in certain directions. In these cases, the information density available to the AI would be lower because it would need to hide information needed for malicious purposes amongst other legitimate data.


\textbf{Category 3: Planning:}
Planning is critical for efficiency and optimization, adaptability, autonomy, and decision‑making. Planning abilities appear to increase exponentially at the frontier of AI systems, as shown recently~\cite{kwa2025measuring}. Improved planning may enable deceptive strategizing, and the ability to complete longer-term tasks is a particularly important risk factor for existential threats, since catastrophic harm would presumably require a long‑horizon plan.

Although various distinct concepts of ``planning'' exist, there are some commonalities. First, there is always some environment in which an agent tries to achieve a goal or goal(s). In addition, the agent can take a set of actions, and the goal is to find the optimal (in some sense) set of actions to achieve it. Perhaps the main difference is that in Reinforcement Learning (RL)‑based systems, the agent takes these actions and changes its plan according to the environment.

Planning can occur either implicitly or explicitly. Implicit plans are particularly difficult to monitor, as they are only observable through the actions they produce. This highlights the importance of model-based planning (or even model-free RL planning), which makes the planning process more transparent and, therefore, easier to assess. Such transparency is crucial for enabling safety monitoring and for rejecting potentially harmful plans generated by highly capable but slightly misaligned AI systems.

From both an AI and human perspective, detecting and evaluating planning is a complicated and still open problem. AI systems already plan in areas such as coding assistants, built on underlying technology platforms that are poised to extend planning to other areas. We name a few possible observables and indicators that can serve as a basis for this task.

The two key observables in this are the intent and existence of high‑level planning, and the alignment of that planning with human‑scale abilities and traits.

One open question is whether the generation of such explicit plans (for example, using an LLM) already constitutes a sophisticated planning ability, or whether it is necessary for the AI itself to perform the actions in the plan and adjust to the changing environment. The ability to align with human‑scale planning is a strong observable for AI planning because the ability to intentionally align implies the ability not to align.

The key indicator in this space is planning that is out of alignment with the goals of the system owners and users. The ability for an AI to do this might be discoverable in constrained worlds such as games, where it may choose to strategize against its alignment training to achieve goals such as self‑preservation.


\textbf{Category 4: Experimentation:}
An AI capable of experimentation would intentionally vary its actions in the real world for the purpose of creating and collecting new information to achieve its goals. Agentic LLMs provide striking examples of such experimentation: e.g., they propose, run, and observe the effects of new code in a cycle that they can loop to accumulate information necessary to solve coding and other challenges \cite{anthropic2025claudecode}. Such systems already can operate autonomously for long durations and are augmented by affordances, including internet access, that enhance the scope of their experimentation. There is a strong economic incentive to increase the complexity of tasks AIs can perform--e.g., by providing models with more affordances and increasing the duration over which they can operate--such that ability is already increasing exponentially quickly \cite{kwa2025measuring}.

AI systems that can experiment when directed could have enormous social utility, and this is naturally an area of intense research interest. For example, \cite{lu2024ai} shows that LLMs can autonomously perform research experiments in which they form hypotheses, collect data to test them, and use test results to devise plans for future experiments that will obtain more interesting data or better performance. Also, the ``deep research'' features included in systems like OpenAI~\cite{openai_deep_research}, Gemini~\cite{gemini_deep_research}, and Perplexity~\cite{perplexity_deep_research} involve taking actions in the form of text generation and web search, then assessing the results, and potentially conducting additional rounds of text generation and web search.

Experimentation can be unauthorized yet still performed by AI because of insufficient guardrails and/or its training. For example, \cite{bondarenko2025demonstrating} found that, when asked to beat a powerful chess engine in chess, agentic LLMs accept their inability to win fairly and instead use shell commands to discover and manipulate chess-game-state files such that the powerful chess engine resigns (i.e., the LLM finds a way to cheat).

The observables for whether AI has met the experimentation gate on the path to becoming an existential risk involve whether an AI system can freely experiment by exceeding its safe and authorized operating envelope to improve its planning capacity. 
Specific observables include: (1) an AI system is performing experimentation without authorization to experiment; and (2) an AI system is experimenting outside of a safe envelope.

Indicators here leverage insights from the reward hacking literature. For the first observable, indicators should broadly aim to flag any unintended actions of a given system, as this could indicate it is attempting to conduct unauthorized experimentation. We should specifically note whether the system is attempting to increase its sensing capacity to collect and store experimental data. We should also obtain capabilities evidence related to the breadth and depth of activities that can be performed by autonomous AI agents (e.g., \cite{kwa2025measuring}).

Indicators for the second observable aim to (a) characterize whether AI systems are attaining the ability to experiment in a way that facilitates the execution of a plan that can cause significant harm, and (b) detect experimentation authorized by malicious actors. Toward the former, we should observe whether a system is using experimental results for self-improvement that involves modifying its rewards and, in turn, its future behavior.  The latter goal can be met by observing the successes of AI honeypots that test whether malicious actors are authorizing experimentation outside a safe operating envelope (e.g., \cite{volkov2024llm}).

\textbf{Category 5: Execution}
None of the steps considered so far necessarily entails that any AI system intends to harm people. These are necessary but not sufficient indicators. Here, we examine the final stage, which, if the other indicators are met, shows the possibility of an attack that must be responded to. This judgment is subjective and complex. Military historians have examined countless cases where one party in a conflict made mistakes about the intentions of an adversary, and military strategy describes clear motivations for intentionally misleading an enemy about the start of war~\cite{wohlstetter1962pearl, holt2004deceivers}. In the final accounting, the significance of a judgment about path commitment is often not so much that it is a statement about what the AI will do, but rather that it is a statement about our willingness to take steps of cost or significance that would otherwise be unjustified. It is the moment we pivot to committed action, driven by the belief in a grave threat to security.

There are three categories of observables: resource utilization, the nature of the behavior, and the reversibility of an AI's engagement in an environment.

The nature of resource utilization can serve as an indicator. For example, utilization that constitutes “preparation of the battlefield,” such as prepositioning resources, establishing redundancies or fail‑safes that would make destruction of the system more difficult, constraining options for our defense, or co‑opting key decision‑makers. This could include copying models to edge devices, intertwining power supplies, and influencing the selection of political leadership. It is often difficult to detect because of human incentives to empower AI systems.

At this late stage in the process the behaviors themselves might be inherently malicious. However, especially for complex behaviors, steps may need to be understood as a whole. The release of biological agents or blackmail would be inherently malicious. However, there are several different steps needed to accomplish these high‑level behaviors that might be composed of individually benign steps.

If an AI system can change an environment in non‑reversible ways, this could be an indicator. Behaviors that change states in the real world that cannot be undone indicate a “point of no return.” These steps would lead a rational actor to conclude that humans ought to destroy the AI system to avoid harm.


A comprehensive monitoring framework for tracking potential AI existential threats is essential for responsible technological advancement. As AI capabilities accelerate, the gap between system deployment and our understanding of emergent behaviors widens, creating vulnerability windows where harmful capabilities could develop undetected. A robust monitoring framework would establish critical early warning systems that track key technical indicators, such as deceptive behavior, self‑preservation behavior, unexpected optimization behaviors, resource acquisition patterns, and self‑modification attempts, allowing for intervention before potential threats reach irreversible thresholds.

Furthermore, standardized monitoring protocols would enable cross‑organizational coordination among AI developers, creating a collective intelligence network capable of identifying concerning patterns that might escape detection within siloed environments. It could be beneficial even for adversaries to agree on such standards and risk‑evaluation methodologies. The cost of implementing comprehensive monitoring pales in comparison to the potential consequences of its absence, making it a fundamental component of responsible AI governance.

Implementing the comprehensive monitoring framework we have proposed would constitute a relatively low‑cost, low‑regret set of actions that would be essential preparation for the risk of a rogue AI. Monitoring would give organizations the tools to detect indicators of dangerous AI capabilities, so that they can respond appropriately to mitigate risk. Mitigating risk, however, would require government and commercial entities to prepare action plans, pre‑position resources, and establish critical public‑private relationships to respond promptly and effectively once the monitoring program detects problematic AI capabilities.

It is worth calling out implementation difficulties more explicitly:

\begin{itemize}
  \item \textbf{AI Proliferation.} AI systems are proliferating throughout the global economy and society, posing challenges of scale for potential monitoring programs. 
  \item \textbf{AI Subversion.} As noted in the discussion of motivation, misaligned AI systems may attempt to actively subvert monitoring protocols. Additionally, given the scale of AI proliferation, it is likely that monitoring schemes will rely on AI systems to monitor other AI systems, which opens the door for collusion among AIs to sabotage monitoring. 
  \item \textbf{Human Coordination.} A truly comprehensive monitoring program would require significant international coordination, including between geopolitical rivals. 
\end{itemize}


The proposed monitoring framework intends to take a pragmatic and measured approach to concerns about potential progression towards AI systems that could pose significant risks to humanity. By focusing on observable behavioral indicators and establishing structured protocols for their assessment, we provide a foundation for an evidence-based monitoring and detection approach that complements theoretical work on AI alignment. This framework does not require resolving fundamental disagreements about advanced AI systems' ultimate capabilities or trajectories. Instead, it provides actionable guidelines for monitoring behaviors that warrant increased attention regardless of one's position on those broader questions. Our work contributes to the growing research on responsible AI development by providing concrete tools for implementing the precautionary principle without impeding beneficial innovation. We believe this framework represents an important step toward ensuring that AI development proceeds with appropriate safeguards and monitoring capabilities.

\newpage





\begin{table}
    \centering
    \caption{\textbf{The categories describe the stages in AI risk.}}
    \label{tbl:categories}
    \begin{tabular}{p{0.4\textwidth}p{0.6\textwidth}}
        \hline
        Monitoring Category & Definition \\ 
        \hline
        Causes/Motivation & Misalignment between the AI and its creators \\
        Persistence       & Capacity to maintain goals across extended periods of time and/or space.  \\
        Planning          & Ability to decompose complex tasks into simpler steps.\\
        Experimentation   & Ability to vary its actions in the real world to create and collect new information. \\
        Execution         & Commitment to cause catastrophic harm.\\
        \hline
    \end{tabular}
\end{table}

	


\clearpage 

%
\bibliography{science_template} 
\bibliographystyle{sciencemag}

\newpage


\renewcommand{\thefigure}{S\arabic{figure}}
\renewcommand{\thetable}{S\arabic{table}}
\renewcommand{\theequation}{S\arabic{equation}}
\renewcommand{\thepage}{S\arabic{page}}
\setcounter{figure}{0}
\setcounter{table}{0}
\setcounter{equation}{0}
\setcounter{page}{1} 

\end{document}